\documentclass[showpacs,aps,graphicx,twocolumn]{revtex4}
\usepackage{graphicx}

\begin{document}

\title{Single-photon entanglement concentration for long-distance quantum communication\footnote{Published in
Quantum Information and Computation \textbf{10} (3\&4), 271-280
(2010).}}

\author{Yu-Bo Sheng,$^{1,2}$ Fu-Guo Deng,$^{3}$\footnote{Author to whom correspondence should be
 addressed.\\
Email address: fgdeng@bnu.edu.cn}  and Hong-Yu Zhou$^{1,2}$}
\address{$^1$ Key Laboratory of Beam Technology and Material
Modification of Ministry of Education, and College of Nuclear
Science and Technology, Beijing Normal University,
Beijing 100875,  China\\
$^2$ Beijing Radiation Center, Beijing 100875,  China\\
$^3$ Department of Physics, Beijing Normal University, Beijing
100875, China }
\date{\today }

\begin{abstract}
We present a single-photon entanglement concentration protocol for
long-distance quantum communication with quantum nondemolition
detector. It is the first concentration protocol for single-photon
entangled states and it dose not require the two parties of quantum
communication to know the accurate information about the coefficient
$\alpha$ and $\beta$ of the less entangled states. Also, it does not
resort to sophisticated single-photon detectors, which makes this
protocol more feasible in current experiments. Moreover, it can be
iterated to get a higher efficiency and yield. All these advantages
maybe make this  protocol have more practical applications in
long-distance quantum communication and quantum internet.
\end{abstract}
\maketitle

\section{introduction}

Quantum entanglement is a unique phenomenon and its distribution
over a long distance is of vital importance in quantum information.
Many quantum processes require entanglement
\cite{book,rmp,densecoding,teleportation}. The simplest entanglement
may be a single-photon entanglement with the form
$\frac{1}{\sqrt{2}}(|1\rangle_{A}|0\rangle_{B}+|0\rangle_{A}|1\rangle_{B})
=\frac{1}{\sqrt{2}}(a^{\dagger}+b^{\dagger})|0\rangle$. It is a
superposition state in location A or B. Here, $\vert 0\rangle$ and
$\vert 1\rangle$ represent the photon number 0 and 1, respectively.
Currently, the most important application for single-photon
entanglement may be the quantum repeater protocol in long-distance
quantum communication \cite{repeater1,repeater2,repeater3}. For
instance, in
 Duan-Lukin-Cirac-Zoller (DLCZ) quantum communication scheme
\cite{repeater1}, with one pair source and one quantum memory at
each location, the quantum repeater is to entangle two remote
locations A and B. The pair sources are coherently excited by
synchronized classical pumping pulses, and then they emit a pair
with a small probability $p/2$, corresponding to the state
\begin{eqnarray}
[1+\sqrt{\frac{p}{2}}(a^{\dagger}a'^{\dagger}+b^{\dagger}b'^{\dagger})+o(p)]|0\rangle.
\end{eqnarray}
Here, $a^+$ ($b^+$, $a'^+$ or $b'^+$) is the creation operation for
the mode $a$ ($b$, $a'$ or $b'$). The two parties of quantum
communication, say the sender Alice and the receiver Bob, store the
modes $a$ and $b$ in their quantum memories in locations A and B,
respectively, and send $a'$ and $b'$ to a station located in the
middle of A and B, where they are combined by a 50:50 beam splitter
(BS). One can create the single-photon entangled state
$\frac{1}{\sqrt{2}}(a^{\dagger}+b^{\dagger})|0\rangle$ between two
distant locations A and B by detecting the photon after the BS.
Furthermore, this entangled state can be converted into a two-atom
entangled state with the form
$\frac{1}{\sqrt{2}}(|e\rangle_{A}|g\rangle_{B}+|g\rangle_{A}|e\rangle_{B})$
\cite{atom,cabrillo,chou1,chou2}. Here $|e\rangle$ is the excited
state and $|g\rangle$ is the grounded state of a two-level atom. In
this way, we can extend the entanglement to two long-distance
locations by using entanglement swapping to complete the task of
long-distance quantum communication \cite{repeater1}.

In a practical quantum repeater, Alice and Bob cannot ensure that
their entangled states are maximal ones. That is, they cannot ensure
the pair sources excited by the synchronized classical pumping
pulses always have the same probability, which means they usually
obtain some pure entangled states, instead of maximally entangled
ones. Meanwhile, in a practical transmission, an entangled quantum
system inevitably interacts with its environment, which will degrade
its entanglement with another form.  For example, a maximally
entangled state may become a mixed entangled one, which  may make a
long-distance quantum communication infeasible. The way of
distilling a set of mixed entangled states into a subset of
maximally entangled states is named as entanglement purification
\cite{Bennett1,Deutsch,Pan1,Simon,shengpra1}. Another method of
distilling a set of less entangled pure states into a subset of
maximally entangled states, which will be detailed here, is named as
entanglement concentration. In 1996, Bennett \emph{et al.} proposed
an entanglement concentration protocol which is called Schmidt
projection method \cite{Bennett2}. Another similar protocol is
called entanglement swapping \cite{swapping1,swapping2}. But the
entanglement swapping protocol  fails to concentrate single-photon
entanglement here. In 2001, Yamamoto \emph{et al.} \cite{Yamamoto}
and Zhao \emph{et al.} \cite{zhao1} independently proposed two
similar two-photon entanglement concentration protocols based on
linear optical elements. In 2008, an efficient two-photon
entanglement concentration protocol based on cross-Kerr nonlinearity
\cite{shengpra2} was presented. In these three two-photon
entanglement concentration protocols
\cite{Yamamoto,zhao1,shengpra2}, it is unnecessary for Alice and Bob
to know the coefficients of the less entangled states accurately.


Although there are some good entanglement purification
\cite{Bennett1,Deutsch,Pan1,Simon,shengpra1} and entanglement
concentration schemes
\cite{Bennett2,swapping1,swapping2,Yamamoto,zhao1,shengpra2}, they
all focus on the polarization entanglement states for photon pairs,
not single-photon entanglements. Fortunately, Sangouard \emph{et
al.} proposed the first single-photon entanglement purification
protocol \cite{singlepurification} with linear optics in 2008. It is
used to purify the phase error. It is rather simple and feasible
with current technology, but it is an important progress for the
implementation of quantum repeater protocols for long-distance
quantum communication. In this paper, we will propose an
entanglement concentration protocol for single-photon entanglement
with quantum nondemolition detectors based on cross-Kerr
nonlinearity. It is the first entanglement concentration protocol
for single-photon entangled states $(\alpha a^+ + \beta b^+)\vert
0\rangle$ and it dose not require the two parties to know the
accurate information about the coefficient $\alpha$ and $\beta$ of
the less entangled pure states. Also, it does not resort to
sophisticated single-photon detectors, which makes this protocol
more feasible in current experiments. Moreover, this protocol can be
iterated to get a higher efficiency and yield than the conventional
entanglement concentration protocols. This single-photon
entanglement concentration protocol and the entanglement
purification protocol in Ref.\cite{singlepurification} may
constitute important progresses for the implementation of the
quantum repeater protocols based on single-photon entanglement in
long-distance quantum communication \cite{repeater1} and quantum
internet.

\section{The problem of entanglement concentration for long-distance quantum communication}

In a practical manipulation, we cannot ensure the pair sources
excited by the synchronized classical pumping pulses always have the
same probability. We still take DLCZ scheme \cite{repeater1} as an
example to describe the principle. The pair source in location A is
coherently excited and can emit a pair with the following form:
\begin{eqnarray}
|0\rangle_{a}|0\rangle_{a'}+\sqrt{\frac{p_{a}}{2}}a^{\dagger}
a'^{\dagger}|0\rangle_{a}|0\rangle_{a'}+o(p_{a}),
\end{eqnarray}
but in location B, the pair source may emit a pair with the form:
\begin{eqnarray}
|0\rangle_{b}|0\rangle_{b'}+\sqrt{\frac{p_{b}}{2}}b^{\dagger}
b'^{\dagger}|0\rangle_{b}|0\rangle_{b'}+o(p_{b}).
\end{eqnarray}
$\frac{p_{a}}{2}$ and $\frac{p_{b}}{2}$ are two different
probabilities for locations A and B, respectively. In this time,
the whole system evolves as:
\begin{eqnarray}
[1+\sqrt{\frac{p_{a}}{2}}a^{\dagger}
a'^{\dagger}|0\rangle_{a}|0\rangle_{a'}+\sqrt{\frac{p_{b}}{2}}b^{\dagger}
b'^{\dagger}|0\rangle_{b}|0\rangle_{b'}+o(p)]|0\rangle.
\end{eqnarray}
Finally, after the detection of the photon combined by a BS, the
single-photon entangled state becomes
$(\sqrt{\frac{p_{a}}{2}}a^{\dagger}+\sqrt{\frac{p_{b}}{2}}e^{i\theta_{AB}}b^{\dagger})|0\rangle$.
We can rewrite it as:
\begin{eqnarray}
|\Psi'\rangle_{ab}=(\alpha a^{\dagger}+\beta
e^{i\theta_{AB}}b^{\dagger})|0\rangle ,\label{state5}
\end{eqnarray}
where $|\alpha|^{2}+|\beta|^{2}=1$. $\theta_{AB}$ is the relative
phase between A and B.  This relative phase is sensitive to  the
path length instabilities between two remote entangled pairs, and it
will make the long-distance quantum communication difficult
\cite{repeater0,repeater}. Eq.(\ref{state5}) is the entanglement of
photonic modes, and we can convert it to the memory modes with
$M_{A}$ and $M_{B}$. The memory modes will  let the two memories be
in the excited state or in the ground state. Meanwhile, the quantum
memory systems can also lead to the same phenomena with different
probabilities for the photons storing in A and B as they both
interact with their environments.

\begin{figure}[!h]
\begin{center}
\includegraphics[width=6.5cm,angle=0]{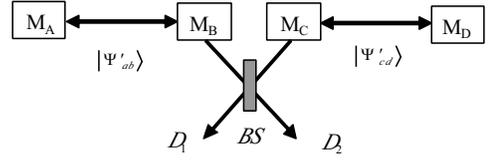}
\caption{The setup of entanglement connection in the DLCZ protocol
\cite{repeater1}. BS is a 50:50 beam splitter. After this swapping,
if one of the detectors registers one photon, the entanglement
between $M_{A}$ and $M_{D}$ can be set up.}\label{connetion}
\end{center}
\end{figure}

After the entanglement generation, we need to extend the
entanglement to long distance with entanglement swapping for
long-distance quantum communication. See from Fig.\ref{connetion},
if the entangled state of $M_{A}$ and $M_{B}$ and that of $M_{C}$
and $M_{D}$ are both the maximally entangled ones, we can easily
establish the maximal entanglement between $M_{A}$ and $M_{D}$
\cite{repeater1}. However, if we cannot get the maximally entangled
states during the stage of entanglement generation, or the
environment noises degrade the maximally entangled states to the
form of Eq.(\ref{state5}), the combination of $|\Psi'\rangle_{ab}$
and $|\Psi'\rangle_{cd}$ can be written as:
\begin{eqnarray}
&&|\Psi'\rangle_{ab}\otimes|\Psi'\rangle_{cd}\nonumber\\
&&=(\alpha a^{\dagger}+\beta
e^{i\theta_{AB}}b^{\dagger})\otimes(\alpha
c^{\dagger}+\beta e^{i\theta_{CD}}d^{\dagger})|0\rangle\nonumber\\
&&=(\alpha^{2}a^{\dagger}c^{\dagger}+\beta^{2}e^{i(\theta_{AB}+\theta_{CD})}b^{\dagger}d^{\dagger})|0\rangle\nonumber\\
&&+(\alpha\beta e^{i\theta_{AB}}
b^{\dagger}c^{\dagger}+\alpha\beta e^{i\theta_{CD}}
a^{\dagger}d^{\dagger})|0\rangle.\label{state6}
\end{eqnarray}
Here we let  $|\Psi'\rangle_{cd}$ have the same form as
$|\Psi'\rangle_{ab}$, i.e., $|\Psi'\rangle_{cd}=(\alpha
c^{\dagger}+\beta e^{i\theta_{CD}}d^{\dagger})|0\rangle$. The BS
will makes
$b^{\dagger}|0\rangle\rightarrow\frac{1}{\sqrt{2}}(D^{\dagger}_{1}+D^{\dagger}_{2})|0\rangle$
and
$c^{\dagger}|0\rangle\rightarrow\frac{1}{\sqrt{2}}(D^{\dagger}_{1}-D^{\dagger}_{2})|0\rangle$.
After the 50:50 BS, from Eq.(\ref{state6}) we find that if one of
the detectors clicks one photon, we get
\begin{eqnarray}
|\Psi''\rangle_{ad}=(\alpha^{2}a^{\dagger}\pm\beta^{2}d^{\dagger}e^{i(\theta_{AB}+\theta_{CD})})|0\rangle.\label{state7}
\end{eqnarray}
The '+' or '-' depends on the click of the detector $D_{1}$ or
$D_{2}$. Compared with Eq.(\ref{state5}), the entanglement of
Eq.(\ref{state7}) is degraded after the entanglement connection.
This problem can be generalized to a more general case. For
instance, we perform entanglement swapping protocol for $n$ times to
connect the entanglement between the remote locations A and K, we
will get
\begin{eqnarray}
|\Psi^{n+1}\rangle_{ak}=(\alpha^{n+1}a^{\dagger}\pm\beta^{n+1}k^{\dagger}e^{i\theta_{AK}})|0\rangle\label{state8}.
\end{eqnarray}
For $\alpha\neq\beta$, the entanglement decreases more and more, and
we will fail to establish a perfect long-distance entanglement
channel for quantum communication. In other words, a long-distance
quantum communication with a practical manipulation of entanglement
generation and a practical transmission requires the entanglement
concentration of single-photon entangled states.

\section{entanglement concentration of single-photon entanglements}

Cross-Kerr nonlinearity has been wildly studied
\cite{QND1,QND2,discriminator,qubit1,qubit2,qubit3}, such as  the
construction of a CNOT gate, the discrimination of unknown qubits,
Bell-state analysis, and so on. The Hamiltonian of the cross-Kerr
nonlinearity is $H_{ck}=\hbar\chi a^{+}_{s}a_{s}a^{+}_{p}a_{p}$
\cite{QND1,QND2,kerr1,kerr2}, where $a^+_s$ and $a^+_p$ are the
creation operations and $a_s$ and $a_p$ are the destruction
operations. When the coherent beam $|\alpha\rangle_{p}$ and a signal
pulse in the Fock state
$|\Psi\rangle_s=c_{0}|0\rangle_{s}+c_{1}|1\rangle_{s}$ interact with
the cross-Kerr nonlinearity, the whole system becomes:
\begin{eqnarray}
U_{ck}|\Psi\rangle_{s}|\alpha\rangle_{p}&=&
e^{iH_{ck}t/\hbar}[c_{0}|0\rangle_{s}+c_{1}
|1\rangle_{s}]|\alpha\rangle_{p} \nonumber\\
&=& c_{0}|0\rangle_{s}|\alpha\rangle_{p}+c_{1}|1\rangle_{s}|
\alpha e^{i\theta}\rangle_{p},
\end{eqnarray}
where $\theta=\chi t$ and $t$ is the interaction time. One can see
that the phase shift of the coherent beam is proportional to the
number of photons in the Fock state. We can use this good feature to
construct a quantum nondemolition detector (QND).

\begin{figure}[!h]
\begin{center}
\includegraphics[width=7cm,angle=0]{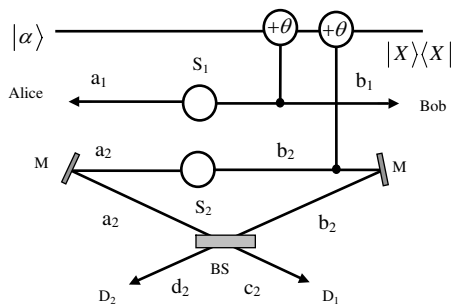}
\caption{The principle of our single-photon entanglement
concentration protocol. Alice and Bob share two single-photon
entangled states with the form
$\frac{1}{\sqrt{2}}(a^{\dagger}+b^{\dagger})|0\rangle$. A 50:50 beam
splitter (BS) is located in the middle of A and B, and it is used to
couple the two modes $a_{2}$ and $b_{2}$. Homodyne detector of the
QND is used to distinguish the photon number. After the detection of
$D_{1}$ and $D_{2}$, the two parties can get some maximally
entangled states with the probability
$2|\alpha\beta|^{2}$.}\label{principle}
\end{center}
\end{figure}

The principle of our single-photon entanglement concentration
protocol is shown in Fig.\ref{principle}. Alice and Bob want to
share the maximally entangled state $|\Psi\rangle_{ab}
=\frac{1}{\sqrt{2}}(a^{\dagger}+b^{\dagger})|0\rangle$. We suppose
that there are two identical less entangled states
$|\Psi\rangle_{a_{1}b_{1}}$ and $|\Psi\rangle_{a_{2}b_{2}}$ shared
by Alice and Bob. One is in the mode ${a_{1}b_{1}}$ and the other
is in the mode ${a_{2}b_{2}}$. The two single-photon less
entangled sates are
\begin{eqnarray}
|\Psi\rangle_{a_{1}b_{1}}&=&(\alpha a^{\dagger}_{1}+\beta e^{i\theta_{A_{1}B_{1}}}b^{\dagger}_{1})|0\rangle,\\
|\Psi\rangle_{a_{2}b_{2}}&=&(\alpha a^{\dagger}_{2}+\beta
e^{i\theta_{A_{2}B_{2}}} b^{\dagger}_{2})|0\rangle.
\end{eqnarray}
In this protocol, we suppose the two sources emit the entangled
state simultaneously, so the path length fluctuations of two
channels $a_{1}b_{1}$ and $a_{2}b_{2}$ can be regarded as the same
one, then the phase $\theta_{A_{1}B_{1}}$  equals to
$\theta_{A_{2}B_{2}}$. We give them a general sign as $\theta_{AB}$.
The original state of these two photons can be rewritten as
\begin{eqnarray}
|\Phi\rangle&=&|\Psi\rangle_{a_{1}b_{1}}\otimes|\Psi\rangle_{a_{2}b_{2}}\nonumber\\
&=& (\alpha^{2}a^{\dagger}_{1}a^{\dagger}_{2}+\alpha\beta
e^{i\theta_{AB}} a^{\dagger}_{1}b^{\dagger}_{2} \nonumber\\
&+&
 \alpha\beta
e^{i\theta_{AB}} a^{\dagger}_{2}b^{\dagger}_{1}
+\beta^{2}e^{2i\theta_{AB}}b^{\dagger}_{1}b^{\dagger}_{2})|0\rangle.
\label{state11}
\end{eqnarray}
$a^{\dagger}_{1}a^{\dagger}_{2}$ and
$b^{\dagger}_{1}b^{\dagger}_{2}$ represent that the two photons both
belong to Alice and Bob, respectively.
$a^{\dagger}_{1}b^{\dagger}_{2}$ and
$a^{\dagger}_{2}b^{\dagger}_{1}$ mean that each of Alice and Bob
 owns a photon. One can see that
$a^{\dagger}_{1}b^{\dagger}_{2}$ and
$a^{\dagger}_{2}b^{\dagger}_{1}$ have the same coefficient
$\alpha\beta e^{i\theta_{AB}}$, and the other two terms have the
different coefficients. Now Bob lets the two modes $b_{1}$ and
$b_{2}$ enter a QND. With a homodyne measurement $\vert
X\rangle\langle X\vert$, Bob may get three different results:
$a^{\dagger}_{1}a^{\dagger}_{2}$ leads no phase shift on the
coherent beam, $b^{\dagger}_{1}b^{\dagger}_{2}$ leads to the phase
shift $2\theta$, and $a^{\dagger}_{1}b^{\dagger}_{2}$ and
$a^{\dagger}_{2}b^{\dagger}_{1}$ lead to the phase shift $\theta$.
So if the phase shift of the homodyne measurement is $\theta$, Bob
requires Alice to keep this result; otherwise both of them  discard
this measurement. In this way, if we omit the global phase factor
$e^{i\theta_{AB}}$, the state of the remaining quantum system
becomes
\begin{eqnarray}
|\Phi\rangle'=\frac{1}{\sqrt{2}}(a^{\dagger}_{1}b^{\dagger}_{2}+a^{\dagger}_{2}b^{\dagger}_{1})|0\rangle.\label{phi22}
\end{eqnarray}
The probability that Alice and Bob get the state $|\Phi\rangle'$ is
$2|\alpha\beta|^{2}$.

The modes $a_{2}$ and $b_{2}$ are reflected and coupled by a 50:50
BS which will make
\begin{eqnarray}
a^{\dagger}_{2}|0\rangle &\rightarrow&
\frac{1}{\sqrt{2}}(c^{\dagger}_{2}-d^{\dagger}_{2})|0\rangle,\\
b^{\dagger}_{2}|0\rangle &\rightarrow&
\frac{1}{\sqrt{2}}(c^{\dagger}_{2}+d^{\dagger}_{2})|0\rangle.
\end{eqnarray}
That is, Eq.(\ref{phi22}) evolves as
\begin{eqnarray}
|\Phi\rangle'&\rightarrow&\frac{1}{2}[a^{\dagger}_{1}(c^{\dagger}_{2}+d^{\dagger}_{2})
+b^{\dagger}_{1}(c^{\dagger}_{2}-d^{\dagger}_{2})]|0\rangle\nonumber\\
&=&\frac{1}{2}[(a^{\dagger}_{1}+b^{\dagger}_{1})c^{\dagger}_{2}+(a^{\dagger}_{1}
-b^{\dagger}_{1})d^{\dagger}_{2}]|0\rangle.
\end{eqnarray}
One can see that if the detector $D_{1}$ fires, the state of the
remaining quantum system is left to be
\begin{eqnarray}
|\Phi_{1}\rangle'=\frac{1}{\sqrt{2}}(a^{\dagger}_{1}+b^{\dagger}_{1})|0\rangle
;\label{statephi1}
\end{eqnarray}
otherwise, the detector $D_{2}$ fires and the quantum system
collapses to
\begin{eqnarray}
|\Phi_{2}\rangle'=\frac{1}{\sqrt{2}}(a^{\dagger}_{1}-b^{\dagger}_{1})|0\rangle
.\label{statephi2}
\end{eqnarray}
They both are maximally single-photon entangled states. There is a
phase difference between Eq.(\ref{statephi2}) and
Eq.(\ref{statephi1}). In polarization entanglement concentration
protocol, one can easily perform a phase-flipping operation to
correct this analogous phase error
\cite{Bennett2,swapping1,swapping2,Yamamoto,zhao1,shengpra2}.
However, for a single-photon entanglement, a phase-flipping
operation is not easily performed. Certainly,  we can convert
Eq.(\ref{statephi2}) into Eq.(\ref{statephi1}) with the help of
quantum memory \cite{repeater1}.

With only one QND, the probability that Alice and Bob get the state
$|\Phi\rangle'$ shown in Eq.(\ref{phi22}) is $2|\alpha\beta|^{2}$.
This protocol does not require Alice and Bob to know the accurate
information about the coefficients $\alpha$ and $\beta$ of the less
entangled pure states. In fact, this probability is not the maximal
one. In the above protocol, Alice and Bob only pick up the instances
that the phase shift is $\theta$ and discard the other cases. If a
suitable cross-Kerr material can be provided, or the interaction
time can be controlled accurately, Alice and Bob will get the phase
shift $\theta=\pi$ when one photon is detected. In this time, if two
photons pass through the QND, they will make the phase shift with
$2\theta=2\pi$. $2\pi$ is the same phase shift as 0. In this case,
Eq.(\ref{state11}) collapses to
\begin{eqnarray}
|\Phi\rangle''=(\alpha^{2}a^{\dagger}_{1}a^{\dagger}_{2}
+\beta^{2}e^{2i\theta_{AB}} b^{\dagger}_{1}
b^{\dagger}_{2})|0\rangle. \label{state19}
\end{eqnarray}
This state is not a maximally entangled one, but with the next step
one can also get the maximally one with single-photon entanglement
concentration. After coupled by the BS, Eq.(\ref{state19}) becomes
\begin{eqnarray}
|\Phi\rangle''&=&[\frac{\alpha^{2}}{\sqrt{2}}a^{\dagger}_{1}(c^{\dagger}_{2}-d^{\dagger}_{2})
+\frac{\beta^{2}e^{2i\theta_{AB}}}{\sqrt{2}}b^{\dagger}_{1}(c^{\dagger}_{2}+d^{\dagger}_{2})]|0\rangle\nonumber\\
&=&[(\frac{\alpha^{2}}{\sqrt{2}}a^{\dagger}_{1}-\frac{\beta^{2}e^{2i\theta_{AB}}}{\sqrt{2}}b^{\dagger}_{1})c^{\dagger}_{2}\nonumber\\
&+&(\frac{\alpha^{2}}{\sqrt{2}}a^{\dagger}_{1}
+\frac{\beta^{2}e^{2i\theta_{AB}}}{\sqrt{2}}b^{\dagger}_{1})d^{\dagger}_{2}]|0\rangle.
\label{state20}
\end{eqnarray}
If the detector $D_{1}$ fires, the state in Eq.(\ref{state20}) is
transformed into
\begin{eqnarray}
|\Phi_{1}\rangle''=(\alpha^{2}a^{\dagger}_{1}-\beta^{2}e^{2i\theta_{AB}}b^{\dagger}_{1})|0\rangle.
\label{state21}
\end{eqnarray}
If the detector $D_{2}$ fires, it is transformed into
\begin{eqnarray}
|\Phi_{2}\rangle''=(\alpha^{2}a^{\dagger}_{1}+\beta^{2}e^{2i\theta_{AB}}b^{\dagger}_{1})|0\rangle.
\label{state22}
\end{eqnarray}

It is easy to find that Eq.(\ref{state22}) has the same form as
Eq.(\ref{state5}). So does the second less entangled state in
Eq.(\ref{state21}). The whole system becomes
\begin{eqnarray}
|\Phi\rangle'''&=&|\Phi_2\rangle''_{a_{1}b_{1}}\otimes|\Phi_2\rangle''_{a_{2}b_{2}} \nonumber\\
&=& (\alpha^{4}a^{\dagger}_{1}a^{\dagger}_{2}+\alpha^{2}\beta^{2}
e^{i2\theta_{AB}} a^{\dagger}_{1}b^{\dagger}_{2} \nonumber\\
&+&
 \alpha^{2}\beta^{2}
e^{i2\theta_{AB}} a^{\dagger}_{2}b^{\dagger}_{1}
+\beta^{4}e^{i4\theta_{AB}}b^{\dagger}_{1}b^{\dagger}_{2})|0\rangle.
\end{eqnarray}
Following  the method above and the help of QND, Alice and Bob pick
up $a^{\dagger}_{1}b^{\dagger}_{2}$ and $
a^{\dagger}_{2}b^{\dagger}_{1}$ with the probability of
$2|\alpha^{2}\beta^{2}|^{2}$, and they keep the other terms for next
iteration. Eq.(\ref{state21}) can also be manipulated with the same
step as that discussed above. Finally, one can perform this protocol
by iteration of the process above and get a higher yield of
maximally entangled states.

\section{discussion and conclusion}

In our protocol, only one QND is used to detect the photon number
in spatial modes. If the modes $a_{1}a_{2}$ and $b_{1}b_{2}$ both
contain one photon, the phase shift of the coherent beam is
$\theta$, which can be easily detected by a homodyne measurement.
Different from single-photon entanglement purification protocol
\cite{singlepurification}, this protocol does not require
sophisticated single-photon detectors as the QND has the function
of a photon number detector.

Let us compare this single-photon entanglement concentration
protocol with the conventional two-photon entanglement concentration
protocols proposed in Refs.\cite{Yamamoto,zhao1,shengpra2}. In fact,
all of these four protocols are based on the principle of Schmidt
projection method, but the conventional protocols are focused on
polarization entanglement states of two-photon quantum systems, not
Fock states of single-photon quantum systems. The polarization beam
splitter \cite{zhao1} and the QND \cite{shengpra2} act as the same
role of parity check. Here the QND acts as the role of a
photon-number detector for two spatial modes, but does not destroy
the Fock states of the two spatial modes. This role can not simply
be replaced with a photon-number detector as a photon-number
detection on the spatial modes $b_1$ and $b_2$ will change their
Fock states and make entanglement concentration fail. Another
difference is that this protocol requires nonlocal operations. In
Refs.\cite{Yamamoto,zhao1,shengpra2}, after the parity check, Alice
and Bob both perform  Hadamard operations on the remaining photons
and detect them locally. However, this single-photon entanglement
concentration protocol works for neighbor nodes in a long-distance
quantum communication. Alice and Bob need to combine the modes
$a_{2}$ and $b_{2}$ into the beam splitter in the middle location of
A and B if the phase shift picks up $\theta$. This combination
principle is similar to the creation of single-photon entanglement
in DLCZ protocol \cite{repeater1}. In principle, this single-photon
entanglement concentration protocol has the efficiency $Y$,
\begin{eqnarray}
Y &=& \sum_{i=1}^{n} Y_i,
\end{eqnarray}
where
\begin{eqnarray}
Y_1 &=& |\alpha\beta|^{2},\nonumber\\
Y_2 &=&
\frac{1}{2}(1-2|\alpha\beta|^2)\frac{|\alpha\beta|^4}{(|\alpha|^4 +
|\beta|^4)^2}\nonumber,\\
Y_3&=&
\frac{1}{2^2}(1-2|\alpha\beta|^2)[1-\frac{|\alpha\beta|^4}{(|\alpha|^4
+ |\beta|^4)^2}]\frac{|\alpha\beta|^8}{(|\alpha|^8 +
|\beta|^8)^2},\nonumber\\
&& \;\;\;\;\;\;\;\;\;\;\;\;\;\;\;\;   \ldots \nonumber\\
Y_n &=& \frac{1}{2^{n-1}}
(1-2|\alpha\beta|^2)\left(\prod^{n-1}_{j=3}[1-\frac{2|\alpha\beta|^{2^{j-1}}}{(|\alpha|^{2^{j-1}}
+
|\beta|^{2^{j-1}})^2}]\right)\nonumber\\
&&\frac{|\alpha\beta|^{2^{n}}}{(|\alpha|^{2^n} + |\beta|^{2^n})^2}.
\end{eqnarray}
This efficiency is the same as that in two-photon entanglement
concentration with QND \cite{shengpra2}, higher than that with PBSs
and photon-number detections \cite{Yamamoto,zhao1} as the efficiency
in the latter is $Y_1$.

Finally, we briefly discuss the problem of  relative phase
instability in DLCZ protocol \cite{repeater1}, which makes the
original DLCZ protocol extremely difficult. The relative phase
between two remote entangled states is caused by the path length
fluctuation, and it must be kept stabilized until the entangled
channel is established \cite{repeater,repeater0}. It exists both in
the entanglement generation stage and the entanglement connection
stage. Here we propose  an entanglement concentration protocol
during the entanglement transformation. We suppose that the two
sources emit the entangled states simultaneously. Therefor, the path
length fluctuation of $a_{1}b_{1}$ and  $a_{2}b_{2}$  should be the
same one. After performing this single-photon entanglement
concentration protocol, the relative phase between two copies of
entangled states is automatically eliminated, and the single-photon
entangled state kept becomes the standard maximally entangled one.
The phase also does not appear in the next entanglement connection
stage.

In conclusion, we have proposed a scheme for single-photon
entanglement concentration, which is realizable with current
technology. This protocol has several advantages. First, it does not
require the two parties of quantum communication to know accurately
the coefficients $\alpha$ and $\beta$ of the single-photon less
entangled pure states $(\alpha a^+ + \beta b^+)\vert 0\rangle$. This
advantage makes this protocol capable of concentrating an arbitrary
less entangled pure state. Second, it does not require sophisticated
single-photon detectors to judge the photon number in each side.
Moreover, this protocol can be iterated to get a higher efficiency
and yield than the ordinary concentration protocols. This
single-photon entanglement concentration protocol and the
entanglement purification protocol in Ref.\cite{singlepurification}
may constitute important progresses for the implementation of the
quantum repeater protocols based on single-photon entanglement.
Furthermore, these two protocols may provide practical applications
in long-distance quantum communication and the construction of the
quantum internet \cite{internet} in future.

\bigskip
\section*{ACKNOWLEDGEMENTS}
This work is supported by the National Natural Science Foundation of
China under Grant No. 10974020, A Foundation for the Author of
National Excellent Doctoral Dissertation of P. R. China under Grant
No. 200723, and  Beijing Natural Science Foundation under Grant No.
1082008.

\end{document}